# Microwave to optical photon conversion by means of travelling-wave magnons in YIG films

M.Kostylev[1] and A.A. Stashkevich[2]

[1]School of Physics, the University of Western Australia, Crawley, 6009 WA, Australia

[2]LSPM (CNRS-UPR 3407), Université Paris 13, Sorbonne Paris Cité, 93430 Villetaneuse, France

*Abstract:* In this work we study theoretically the efficiency of a travelling magnon based microwave to optical photon converter for applications in Quantum Information (QI). The converter employs an epitaxially grown yttrium iron garnet (YIG) film as the medium for propagation of travelling magnons (spin waves). The conversion is achieved through coupling of magnons to guided optical modes of the film. The total microwave to optical photon conversion efficiency is found to be larger than in a similar process employing a YIG sphere by at least 4 orders of magnitude. By creating an optical resonator of a large length from the film (such that the traveling magnon decays before forming a standing wave over the resonator length) one will be able to further increase the efficiency by several orders of magnitude, potentially reaching a value similar to achieved with opto-mechanical resonators.

Also, as a spin-off result, it is shown that isolation of more that 20 dB with direct insertion losses about 5 dBm can be achieved with YIG film based microwave isolators for applications in Quantum Information.

An important advantage of the suggested concept of the QI devices based on travelling spin waves is a perfectly planar geometry and a possibility of implementing the device as a hybrid opto-microwave chip.

## I. Introduction

The magnon-based microwave-light converter is very attractive from the viewpoint of enlarging the potential of the superconducting qubits [1-5]. Such device whose first conversion stage is based on coherent coupling between a ferromagnetic magnon and a superconducting qubit is expected to have a bandwidth of around 1 MHz and thus operates faster than the lifetime of a superconducting qubit currently available (around 100 μs [1]). Moreover, the ferromagnetic magnon coherent coupling to a superconducting qubit has recently been confirmed experimentally [2,3]. The most natural and direct way of up-converting, at the second stage, thus excited microwave magnons (typical frequencies lying in the lower part of the GHz range) to the optical domain is via magneto-optical interactions, thus realizing coherent connection between distant superconducting qubits. In the basic configuration studied so far the microwave magnons, known as magnetostatic Kittel mode are excited in an uniformly magnetized YIG sphere with a diameter of about 1 mm. This mode is spatially homogenuous and can be considered as the lowest ferromagnetic resonance (FMR) localized within a spherical ferromagnetic resonator. The FMR line-width $\Delta H$ is its major parameter determining the quality factor. To improve the efficiency of the magnon-photon magneto-optical (MO) coupling, the latter should be kept as large as possible. This implies a very narrow FMR line which in its turn corresponds to a narrow frequency bandwidth of effective MO interaction $\Delta f_{FMR}$. Thus, the value of FWHM $2\Delta H=0.3$ Oe typical for high-quality YIG specimen corresponds to a very modest bandwidth of 1 MHz.

One has to note that the currently achieved quantum efficiency of the microwave to optical photon coupling via magnons in a YIG sphere is also quite modest $10^{-10}$ [4]. This is much smaller than already achieved with other methods [6-10]. In particular, a 10% coupling was demonstrates through an opto-mechanical route [6], and electro-optics allows one to reach 0.1% [7].

Therefore, in this paper we investigate theoretically an alternative solution to the problem of magnon-to-light conversion. More specifically, we consider travelling magnons in the form of magnetostatic spin waves as a candidate for this role. A spin wave is a wave of electromagnetic nature which is propagating in a saturated ferrimagnetic film and which is strongly retarded (1000 - 10000 times with respect to the velocity of a normal electromagnetic wave) due to a strong resonance interaction with the matter. As a consequence of the slow propagation, the magnetic component of the wave is extremely strong. This is why magnetostatic spin waves can be exceptionally effective in MO modulating devices based on the Faraday Effect.

Further, such a microwave to optical photon converter has an intrinsically planar geometry. Therefore, contrary to the YIG sphere in a microwave cavity concept [4], it potentially allows integration into a hybrid opto-microwave chip of reasonable sizes and fabrication of the microwave-photon to magnon transducers using optical lithography.

First experiments on MO diffraction of light by magnetostatic surface spin waves (MSSW) date back to the early 1970s [11-12]. However, the interaction efficiency was low due to the inadequately chosen configuration; the light was impinging on a YIG film normally. Thus, the interaction length was equal to the film thickness. A major breakthrough was achieved a decade later through introduction of the so-called waveguide MO interaction [13] in which case both the MSSW and the optical wave propagate simultaneously in the YIG film as in a waveguiding structure. In this geometry the interaction length is determined by the MSSW free propagation path, which typically is on the order of several millimeters. As a result, the interaction intensity was increased several orders of magnitude. In the course of investigations that followed in the late 1980s, several basic SW-optical device geometries have been studied, pure YIG films typically having been used as a medium for magneto-optical interaction [14-16]. These experiments have stressed the significance of the improvement of such an important parameter as the interaction efficiency. Not surprisingly, theoretical effort was focused on the specificity of strong MSSW-Light interactions involving multiple-scattering mechanisms depicted mathematically by Feynman-type multiple-scattering diagrams [17].

Further improvement of the interaction intensity (20 – 50 times) was due to the progress in the technology of highly bismuth-doped YIG films with higher values of the Faraday constant (5 – 10 times) [18-20]. Also, a more sophisticated optical geometries were suggested by theoreticians – the focus has been on using bi-layer ferromagnetic films and confining the optical modes in the film plane in order to improve the structure of optical modes and the light/spin wave interaction efficiency [21,22]. All these experiments have been performed at room temperatures and microwave and optical power levels well above the single photon/magnon one. No experiments have been carried out so far at cryogenic temperatures and at the single optical-photon / magnon power level. On the other hand, the first experiments probing microwave properties of magnons in YIG films at millikelvin temperatures have been already carried out [23,24]. Therefore, it is very timely to look at this problem from the theoretical prospective.

The latest theoretical effort [21,22] in this field was focused on telecom applications, therefore its goal was to achieve a 100% conversion of incident *light* into scattered *light* by employing a high-power spin wave signal. The incident spin wave power was of no importance in that case. Furthermore, the efficiency of the conversion of the microwave signal into spin waves by microwave stripline transducers was not included in the theory.

In the present work we focus on a qualitatively different situation. We assume that the microwave signal incident onto the device input port is at a single-photon level. Therefore, a 100% conversion of the incident *light* into an output optical signal is impossible. And we are not interested in achieving the 100% light-to light conversion. Instead, we are interested in a 100% conversion of the weak input *microwave* signal into the output optical one, expressed in a number of generated optical photons per one input microwave photon. The incident optical power can be arbitrary in this situation. We investigate this configuration theoretically and make predictions about the rate of scattering of the guided light from single traveling magnons. In the process of the calculation, the efficiency of microwave photon to magnon conversion is taken into account in full.

In contrast to the case of an YIG sphere in a microwave cavity, theoretical description of coupling of traveling spin waves to microwave photons in microwave stripline transmission lines is much more involved (see e.g. [25]). Furthermore, it is difficult to obtain efficient microwave photon to magnon coupling experimentally, unless the coupling geometry is carefully optimized (which is feasible even in a mass-production setting). Therefore, an extended part of the present paper (Section IIA) is devoted to theoretical treatment of spin wave excitation by stripline antennas. Section IIB presents theoretical details of coupling of travelling-wave magnons to guided optical photons in the YIG film. Section IIIA reports on the results of numerical calculations of the microwave to optical photon conversion efficiency employing the formalism presented in Section II. Ways to further improve the conversion efficiency are discussed in Section IIIB. Conclusions are contained in Section IV.

As a side result of this calculation, we also evaluate the efficiency of a microwave isolator based on travelling Damon-Eshbach magnetostatic surface spin waves [26] in YIG films. It has been recently shown that these devices are very important for isolation of qubits from the noise in the microwave circuits to which they are connected [27]. Previous calculations and experiments showed that losses inserted by a travelling spin wave device in the forward direction can be small – about 5 dB theoretically and 10 dB experimentally (see Figs. 2-4 in Ref. [28]). In our calculation, we evaluate transmission of the device in both directions and find that a YIG-film based isolator employing asymmetric coplanar transducers has very good performance characteristics.

## II. Theory

### A. Excitation and propagation of spin waves in the YIG film waveguide.

In this work we will be dealing with rather thick YIG films, that is why the Damon-Eshbach exchange-free approximation [26] is appropriate. However, if needed the exchange interaction can be easily included in the theory by using the approach from [26].

The geometry of the problem is shown in Fig. 1. It represents a YIG film of thickness $L$ in the direction $y$. In this section, we consider the film being continuous in both in-plane directions $x$ and $z$. The film is placed on top of a microwave stripline. The stripline runs along the $z$-axis and consists of a number of parallel conductors (lines). We will be interested in a general case of a backed coplanar line shown in Fig. 1. The line consists of three parallel conductors of widths $w_1$, $w$, and $w_2$ in the direction $x$. The central conductor of a width $w$ is called the signal line. The two lines of widths $w_1$ and $w_2$ are ground lines. They are separated from the signal line by gaps of widths $\Delta_1$ and $\Delta_2$ respectively. If one of the ground lines is absent ($w_2=\Delta_2=0$), one deals with a microwave transmission line called the asymmetric coplanar line. If both ground lines are absent ($w_1=\Delta_1=w_2=\Delta_2=0$), but the ground plane at $y=-d$ is still present one deals with a microstrip line (We will use an asymmetric coplanar line in examples which will be considered in the discussion sections of the paper.)

The stripline length in the direction $z$ is assumed to be infinite at the first stage of the problem solution. Also, the YIG film will be considered as infinite in both in-plane directions $x$ and $z$ at the first stage. At the second stage, we will assign specific values to the stripline length and the film width in the direction $z$, once an expression for stripline's linear impedance has been obtained. The thickness of the microstrip in the direction $y$ is assumed to be zero. All these assumptions significantly simplify the calculations without any significant loss of generality [25,30,31].

The stripline is supported by a dielectric substrate of thickness $d$ whose other surface (located at $y=-d$) is metalized thus forming a backed coplanar line. The dielectric permittivity of the substrate is $\varepsilon$. A microwave current $I$ flows through the line. The linear density of this current is $j(x)$ may be non-uniform across the widths of both signal and ground lines and obeys the relation as follows:

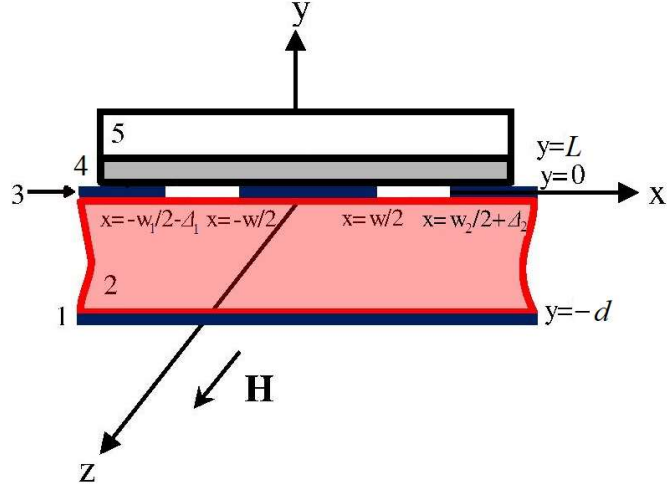

Fig. 1. Vertical cross-section of the geometry of the problem. 1: Stripline ground plane. 2. Stripline substrate. 3 Stripline (ground, signal and ground lines of widths $w_1$, $w$ and $w_2$ respectively). 4. Yttrium iron garnet (YIG) film. 5: Film substrate made of gadolinium gallium garnet. The static magnetic field **H** is applied along the *z*-axis.

$$\int_{-w/2}^{w/2} j(x)dx = I \qquad (1).$$

A microwave Oersted field of the current drives magnetisation precession in the film. The total width of the stipline $w_1+\Delta_1+w+\Delta_2+w_2$ is assumed to be much smaller that the free propagation path of the magnetostatic surface spin waves in epitaxial YIG films (typically several mm for $4^+$ micron thick YIG films). Therefore the excited magnetization oscillation propagates as a plane wave from the stripline transducer in both directions along the *x*-axis. We will term this stripline "the input stripline transducer".

We also assume that a magnetic field $\mathbf{H} = H\mathbf{u}_z$ is applied along the axis *z* thus forming conditions for propagation of a Damon-Eshbach surface wave (MSSW) [26] along the *x*-direction (here $\mathbf{u}_z$ is the unit vector along *z*). Also, an optical guided mode can propagate in the YIG film along the *x*-direction with its evanescent field penetrating into the gadolinium gallium garnet (GGG) substrate of the film. We will deal with optical properties of this system in Section II B, but now let us focus on the formalism of excitation of the spin waves in the film by the input transducer and propagation of the excited spin waves in the film.

The goal of this section is to express the efficiency of spin wave excitation in terms of a number of excited magnons per one microwave photon incident onto the input port of the input transducer. We will use a classical formalism for spin waves for the calculation; the quantum theoretical notions of the numbers of photons and magnons will appear at the very last step of this derivation (Eq.(21)). The reader not interested in the details of this calculation can skip directly to this equation.

The most comprehensive theory of the efficiency of spin wave excitation in this geometry was developed by Vugalter and Gilinski [31]. Our formalism will be similar to the one suggested by them, but because we are interested in maximising the efficiency of all involved interactions, we will be paying greater attention to detail. In Ref.[31] the focus was on obtaining analytical solutions. This was achieved by introducing a number of approximations valid in particular limiting cases, for instance, large values of *d* or small values of *w* with respect to spin wave wavelength. We will keep the theory as general as possible, keeping in mind that derived equations will be solved numerically at the last stage of the calculation, in order to obtain results which are as rigorous as possible. This is because the focus of the present paper is not on derivation of equations but on getting numbers which reflect efficiency of microwave to optical photon conversion via the travelling-magnon route.

Our analysis will be similar to one we used in Refs. [31,26]. The translational invariance of the geometry in Fig. 1 in the *z*-direction enables calculation of a quantity called the complex impedance of stripline. This is achieved by using a quasi-static approach for the description of microwave transmission lines whose central point is an assumption that all the fields can be considered as uniform in the direction *z*. This allows decomposing the problem to a two dimensional one, where all dynamic variables depend only on *x* and *y* [25,30,31].

The magnetization dynamics in the films are described by the linearized Landau-Lifshitz equation

$$\frac{\partial \mathbf{m}}{\partial t} = -|\gamma|(\mathbf{m} \times \mathbf{H} + \mathbf{M} \times \mathbf{h}), \quad (2)$$

where the static magnetization within the film is given by $\mathbf{M} = M\mathbf{u}_z$ where *M* is the saturation magnetization for the film. The dynamic magnetization vector **m** has only two components (perpendicular to **M**) and can be represented as $m_x\mathbf{e}_x + m_y\mathbf{e}_y$. The dynamic dipole field of precessing magnetization **h** is attained from solving Maxwell's equations in the magnetostatic approximation

$$\nabla \times \mathbf{h} = 0, \quad (3)$$

$$\nabla \cdot \mathbf{h} = -\nabla \cdot \mathbf{m}, \quad (4)$$

$$\nabla \times \mathbf{e} = -i\omega\mu_0(\mathbf{h} + \mathbf{m}), \quad (5)$$

where **e** is the microwave electric field generated by the dynamic magnetization due to Faraday induction. In the framework of the quasi-static approach for the stripline description, from the symmetry of the problem it follows that **e** has only a z-component: $\mathbf{e} = e_z\mathbf{u}_z$. A solution for all dynamic variables entering the equation – **m**, **h** and **e** – is obtained representing it as a set of plane waves propagating along *x*:

$$\mathbf{m}, \mathbf{h}, \mathbf{e} = \int_{-\infty}^{\infty} \mathbf{m}_k, \mathbf{h}_k, \mathbf{e}_k \exp(i\omega t - ikx) dk, \quad (6)$$

where $\omega$ is the angular frequency of the microwave current flowing through the input transducer and $k$ is the Fourier wave number.

The analytical expressions for $\mathbf{m}_k$, and $\mathbf{h}_k$ are given in Appendix A. Once these expressions have been derived, one can calculate $\mathbf{m}(x)$, $\mathbf{h}(x)$ and $\mathbf{e}(x)$ by carrying out the inverse Fourier transformation (6). It this work, the transformation is carried out numerically. In order to enable this, magnetic losses $\Delta H$ in the material are introduced into the expressions. We do this by replacing $H$ with $H+i\Delta H$ in the final expressions [32]. Also, a simple analytical solution exists for the far zone of the "spin wave antenna" which the input transducer actually represents.

We will return to the far-zone solution below. Let us now turn to the expression for the complex linear impedance of the transducer $Z_r$ (often called the "Radiation impedance of a spin wave antenna" [31]). It may be obtained in the framework of the "Induced Electromotive force" method, as suggested in [30]. Following this method,

$$Z_r = -\frac{1}{I^2} \int_{-w/2}^{w/2} \overline{j}(x) e_z(x, y=0) dx, \quad (7)$$

where $I$ is the microwave current through the antenna, $j(x)$ is its linear density and $e_z(x, y=0)$ is the $z$-component of the microwave electric field at the level of the antenna ($y=0$) and the dash above $j$ denotes complex conjugation. This electric field has two components – the self-field of the microstrip line and the electric field of the precessing magnetization in the film. Outside the spin wave frequency band the latter field vanishes, and $e_z(x, y=0)$ reduces to the self-field of the stripline [31,33]. This property will be used in the following in order to calculate the characteristic impedance and the complex propagation constant for the transducer.

Starting with this expression, a solution for $Z_r$ is obtained. We do not derive this solution here because it is a simplified version of a more general result from [29] and also because very similar solutions exist in the literature [25,30,31,33]. The final formula for the complex radiation impedance has the form as follows

$$Z_r = -\frac{2\pi}{|I|^2} \int_{-\infty}^{\infty} \overline{j}_k e_{zk}(y=0) dk, \quad (8)$$

where the Fourier transform of the electric field at $y=0$ reads

$$e_{zk}(y=0) = -\frac{i\omega\mu_0}{|k|}\left(\cosh(|k|d) j_k + \tanh(|k|d) h_x(y=0)\right). \quad (9)$$

In order to find the Fourier transform $j_k$ of the microwave current density in the transducer, we may use the fact that in the quasistatic approximation, the continuity equation for $j(x)$ [31] reduces to

$$j(x) = -i(\omega/\gamma_c)\rho(x), \quad (10)$$

where $\rho(x)$ is the linear charge density distribution across the transducer width for the current and $\gamma_c$ is the complex propagation constant for the stripline. This implies that $j_k$ scales as the Fourier transform of the charge density $\rho_k$. $j_k = -i(\omega/\gamma_c)\rho_k$, where the concrete value of the constant $i(\omega/\gamma_c) = K$ is of no importance for the following. The charge density can be obtained by solving the electrostatic problem for the transducer [29]. Because in the following we will be dealing with transducers of an unusual shape, in the present work we find $\rho(x)$ self-consistently.

To formulate the self-consistent problem, we use the fact that in the electrostatic approximation, the electric potential $u(x)$ should be uniform across the widths (in the direction $x$) of the transducer electrodes (strips). Requiring this and using Eq.(B11) from [29] which relates the Fourier transform of the potential $u_k$ to the Fourier transform of the charge density $\rho_k$ one arrives at an integral equation for $\rho(x)$. This equation can be easily solved numerically to yield $\rho(x)$ for a stripline geometry of interest. $\rho(x)$ is then Fourier transformed numerically to yield $\rho_k$. Then $j_k = K\rho_k$. This method takes into account the dielectric properties of the ferromagnetic film and its substrate while calculating $j_k$ and $\rho_k$, but it does not take into account the magnetic properties of the film. However, in [28] it has been shown that the influence of the excited spin waves on the $x$-dependence of $j$ is negligible for most of the spin wave vector range, except the upper edge of the range. In the present work, we will be interested in smaller spin wave wavenumbers, therefore this approximation is appropriate.

Once we have obtained $j_k$, the complex linear impedance $Z_r$ is easily computed numerically. ($K$ cancels out in the course of this derivation, because of the static nature of the involved fields.) $Z_r$ is then transformed into the input impedance for the stripline by utilizing established formulas [33,29]. To maximize the efficiency of spin wave excitation, one has to maximize the microwave current through the stripline. This is obtained by shorting the end of the antenna. In order to incorporate this requirement in our model, now we assume that the stripline has a finite length $l_a$ (in the direction $z$) and the width of the film in the direction $z$ is also finite and equals the same $l_a$.

The input impedance for a stripline obeys Telegrapher Equations [34]. For a stripline of a length $l_a$ whose other end is shorted one has

$$Z_{in} = Z_c \tanh(\gamma_c l_a), \quad (11)$$

where $Z_c$ is the characteristic impedance for the stripline. Both $Z_c$ and $\gamma_c$ relate to the linear parallel capacitive conductance $i\omega C$ and in-series inductive $Z_L$ reactance of the stripline. The linear capacitance $C$ is obtained as a by-product of the solution of the electrostatic problem for $\rho(x)$ (see Eq.(B9) in [29]), and $Z_r$ plays the role of $Z_L$ in our case, as it reduces to $Z_L$ outside the spin wave band (see the comment after Eq.(7) above). Hence,

$$Z_c = \sqrt{Z_r/(i\omega C)}, \quad (12)$$

$$\gamma_c = \sqrt{i\omega C Z_r}. \quad (13)$$

Then the complex reflection coefficient from the input transducer input port reads:

$$\Gamma = (Z_{in} - Z_0)/(Z_{in} + Z_0), \quad (14)$$

where $Z_0 = 50\Omega$ is the characteristic impedance of the microwave feeding line.

Let us now assume that the microwave voltage $V_{in}$ incident onto the input port of the transducer has an amplitude of 1 Volt. The theory of transmission lines then allows us to express the current in the transducer as a function of the position $0 \leq z \leq l_a$ ($z=0$ coincides with the transducer input port and $z = l_a$ with the shorted antenna end):

$$I(z) = \frac{V_{in}}{Z_c} \frac{1-\Gamma}{1-\Gamma \exp(-2\gamma l_s)} \left[ \exp(-\gamma z) - \exp(-\gamma l_s + \gamma z) \right]. \quad (15)$$

As follows from Eq.(15), the amplitude of the driven precession of magnetization below the input transducer will be a function of the co-ordinate $z$ along the transducer. So, each point $z$ will be a point source of a travelling spin wave whose complex amplitude scales as $I(z)$, and the wave front in the far field of antenna will be a result of interferences of these partial waves. Hence, strictly speaking, the problem of formation of the wave front of a travelling spin wave in the far zone of the antenna is 2-dimensional (see e.g. [31]). Solving the two-dimensional problem is beyond the scope of the current work, as it involves consideration of caustic beams which spin waves in in-plane magnetized films are prone to form (see e.g. [35]) if the films are not confined in the z direction. However, if they are confined, a guided spin wave mode will be formed. The **m**(z) distribution for the guided spin wave is not perfectly uniform [36] across the width of the waveguide, however we may neglect this non-uniformity as it is not of central importance for the present paper. Accordingly, below we will assume that the **m**(z) is uniform across the film width (the width coincides with the antenna width $l_a$) everywhere in the far zone of the antenna (the latter also implies that we assume that the guided mode is formed straight after the wave escapes the near antenna zone.)

The energy conservation law tells us that the spin wave energy adjusted for energy losses due to the intrinsic magnetic damping in the film should be the same for any waveguide cross-section. Therefore, we may introduce an effective z-uniform current through the input antenna which produces the same spin wave energy carried through a waveguide cross-section $L \times l_a$. The magnitude of the effective current reads

$$|I_{eff}| = \sqrt{\frac{1}{l_a}\int_0^{l_a}|I(z)|^2 dz}. \quad (16)$$

Note that this definition leaves the phase of the effective current undefined.

The dynamic magnetisation of the travelling spin wave at any position *x* is given by Eq.(6), where **m**$_k$ is given by Eqs. (A1-A2) from Appendix A. In the closed form, the Fourier component of the dynamic magnetisation (for the input microwave signal of 1 Volt, see above) reads:

$$m_{kx(y)}(x,y) = -\frac{2\pi}{\bar{I}_{eff}}\frac{AC_{mx(y)}(|k|)\exp(|k|y) + BC_{mx(y)}(-|k|)\exp(-|k|y)}{\det(\hat{W}(k))}j_k, \quad (17a)$$

where $C_{mx}=1$ and $C_{my}$ is given by Eq.(A3) in Appendix A.

The inverse Fourier transformation of (Eq.(17a)) can be carried out analytically. The analytical solution of the integral in (6) is expressed in terms of exponential integral functions and two complex exponential functions [37]. The exponential integral term describes the near field of the transducer [38] and the complex exponential ones two travelling spin waves propagating in the two opposite directions from the transducer. A similar analytical solution can be obtained for the integral in Eq.(8). The near-field term of that solution yields the antenna linear reactance and the real-valued term represents the spin wave radiation resistance of the antenna. We will not evaluate analytically $Z_r$ in the present work. Instead we will use numerical integration in order to calculate $Z_r$, which is just easier and more accurate.

The near field of an MSSW transducer is localised in the closest vicinity of the transducer, as we previously demonstrated experimentally in Ref.[38]. Outside this area, only the travelling wave contributions to the integral exist. Evaluation of the integral in (6) in this far zone of the transducer using the Residues Theory is straightforward and results in a simple expression as follows

$$m_{x(y)}(x,y) =$$
$$\frac{(2\pi)^3}{\bar{I}_{eff}}\frac{AC_{mx(y)}(|k_0|)\exp(|k_0|y) + BC_{mx(y)}(-|k_0|)\exp(-|k_0|y)}{\frac{d}{dk}(\det(\hat{W}(k_0)))}j_{k_0}\exp(-ik_0 x)\exp(-vx) \quad (17b)$$

where $\det(\hat{W}(k))$ denotes the determinant of the matrix $\hat{W}(k)$ shown in Appendix A and A, B, $C_{mx}$ and $C_{my}$ are coefficients also given in Appendix A. $k_0$ is the value of spin-wave wave number which satisfies the implicit dispersion relation for spin waves $\det(\hat{W}(k)) = 0$. In the limiting case $d=\infty$ it reduces to the Damon-Eschbach dispersion relation [26] for spin waves in a film not sitting on top of a backed stripline:

$$\omega^2(k)/\gamma^2 = H(H+M) + \frac{M^2}{4}[1-\exp(-2kL)].$$

In the following we will refer to the case $d=\infty$ as "a film in vacuum".

The quantity $j_{k_0}$ is obtained as the Fourier transform of $j(x) = const \cdot \rho(x)$, where the value of the constant is given by the condition $\int_{-w_2/2}^{w_2/2} j(x)dx = I_{eff}$.

There are two solutions to the dispersion relation $\det(\hat{W}(k)) = 0$, one for a positive wave number $k$ and one for a negative one. The magnitudes of the two wave numbers are different, as the presence of the metal ground plane of a stripline at $y=-d$ makes the spin wave spectrum non-reciprocal.

The spatial decay factor $\nu$ for the wave is given by the expression as follows:

$$\nu = \frac{d\left[\det\left(\hat{W}(k_0)\right)\right]/dk}{d\left[\det\left(\hat{W}(k_0)\right)\right]/dH}\Delta H, \qquad (18)$$

Where $\Delta H$ is the loss parameter for the film, and, as in Eq.(17b), the argument $k_0$ means that the derivatives are evaluated for $k=k_0$.

The analytical expression (17b) is in excellent agreement with the direct numerical evaluation of the integral in (17a) in the far zone of the transducer. This excellent agreement allows us to use the same analytical approach in order to evaluate the Poynting vector for spin waves [31]. An expression for the Poynting vector for a somewhat simpler geometry of a film in vacuum is shown in Appendix B. The derivation of an expression for the Poynting vector for the geometry of a film on top of a backed coplanar line is also straightforward (not shown here). Numerical evaluation of this expression demonstrates excellent agreement with the expression for the power irradiated by the transducer in the form of spin waves.

The microwave power incident onto the input port of the transducer reads

$$P_{in} = \frac{V_{in}^2}{2Z_0}. \qquad (19)$$

Then from the telegraphist equations it follows that the incident microwave power converted into the magnon power is given by

$$P = \frac{1}{2}\frac{V_{in}^2}{Z_0}\mathrm{Re}\left[(1+\Gamma)(1-\bar{\Gamma})\right].$$

This power is redistributed between the two waves excited by the transducer in the two opposite propagation directions $+k$ and $-k$. Excitation of the Damon-Eshbach spin waves is highly non-reciprocal (see e.g. [39]). The portion of the power $\eta_{nr}$ carried by the spin wave propagating in the positive direction of the axis $x$ (i.e. in the $+k$ direction) is given by [30]

$$\eta_{nr} = R^+ / \mathrm{Re}(Z_r),$$

where $R+$ is the component of the total radiation resistance associated with excitation of spin waves travelling in the $+k$ direction:

$$R^+ = -\frac{2\pi}{|I|^2} \mathrm{Re}\left[\int_0^\infty \bar{j}_k\, e_{zk}(y=0) dk\right].$$

Hence, the power $P^+$ carried by spin waves in the $+k$ direction is given by

$$P^+(x) = \frac{1}{2}\frac{V_{in}^2}{Z_0} \mathrm{Re}\left[(1+\Gamma)(1-\bar{\Gamma})\right] \eta_{nr} \exp(-2\nu x). \qquad (20)$$

A result of numerical evaluation of this expression in the far zone of an asymmetric coplanar transducer ($x > w_2 + \Delta_2 + w_3$) is in excellent agreement with the value of the Poynting vector for spin waves, as has already been stated above.

Eq.(20) also yields an expression for the efficiency of the microwave photon to magnon conversion

$$N_m = \frac{P^+(x=0)}{P_{in}}. \qquad (21)$$

Since the incident microwave photons and the generated magnons have the same frequency (energy), the efficiency of the photon-magnon transduction $N_m$ actually represents the number of generated magnons per one incident photon.

### B. Scattering of optical photons from magnons

In this section we consider two guided optical modes propagating in the YIG film. One is incident, the other is generated due to a photon-magnon interaction. An incident optical mode scatters from a spin wave whose wave vector is collinear to light. This creates the second optical mode which we will call "scattered light". The physics behind the process of the interaction of the guided light with the spin wave is light scattering from a diffraction grid formed by a spatial harmonic modulation of the material's refractive index along the light propagation path. The modulation originates from the dynamic magnetisation of the spin wave – the spatial variation of the magnetisation vector leads to spatially periodic modulation of the strength of the Faraday and Cotton-Mouton Effects. The "diffraction grid" moves with the phase velocity of the spin wave which leads to important peculiarities of light scattering from it. They will be discussed in the very end of the Discussion section.

From the point of view of wave interactions, the magnon-optical photon interaction is a three-wave process. Classically, this process is described by a theory of coupled modes (see Eqs. (16-17) in [21]). The mode coupling coefficient scales as the overlap intergral for the three waves (which reflects spatial correlation of the interacting waves)

$$I_{overlap} = \iiint_{film} \mathbf{E}^{(s)}(x,y,z) \cdot \mathbf{m}(x,y,z) \cdot \mathbf{E}^{(i)}(x,y,z) dV. \qquad (22)$$

Here functions $\mathbf{E}^{(i)}(x,y,z)$, $\mathbf{E}^{(s)}(x,y,z)$, $\mathbf{m}(x,y,z)$ describe the spatial distribution of the electric field in the incident and scattered optical waves and that of the magnetization in the scattering spin wave.

Secondly, the efficiency of the MO interactions depends on the "correlation" of their polarizations (the vector factor). Mathematically, the symmetry of MO coupling in an optically isotropic media is described by the totally antisymmetric Levi-Civita tensor [40]. As a result, the vector factor is expressed via the mixed product of the interacting waves

$$I_{vec} = \left(\vec{e}^{(s)} \cdot \left(\mathbf{m} \times \vec{e}^{(i)}\right)\right) = \sum_{i=x,y,z}\sum_{j=x,y,z}\sum_{l=x,y,z} \delta_{ijl} e_i^{(s)} m_j e_l^{(i)}. \qquad (23)$$

Correspondingly, their polarizations are given by unit vectors $\vec{e}^{(i)}$, $\vec{e}^{(s)}$, $\mathbf{m}$. An interested reader can find all details in an exhaustive theoretical analysis presented in Ref. [41].

A YIG film can be regarded as a planar dielectric waveguide. Due to its specific symmetry such a waveguide supports two types of electromagnetic waves, namely TE-modes $\mathbf{E}^{(n)TE}(x,y) = E_z^{(n)TE}(y)\exp(-i\beta^{(n)TE}x)\vec{e}_z$ and TM-modes $\mathbf{E}^{(n)TM}(x,y) = \left(E_x^{(n)TM}(y)\vec{e}_x + E_y^{(n)TM}(y)\vec{e}_y\right) \cdot \exp(-i\beta^{(n)TM}x)$ with 1<n<N. The value of N, for a given optical wavelength, depends on the film thickness. Here $\beta^{(n)TE}$ and $\beta^{(n)TM}$ are waveguide wavenumbers of respectively *n*-th TE and *n*-th TM mode. The orthogonality of the modes in the sense of the cross-power allows expressing an arbitrary given field distribution as a superposition of waveguide modes. In this regard, the spin wave possesses the structure of a TE-mode with the dynamic magnetization in the *"xy"* plane.

The selection rules relying directly on Eqs.(22,23) boil down to the following. First, due to Eq.(23), interactions between modes of the same type are impossible. Similarly, the overlap integrals in Eq.(22) forbid the interaction if $n \neq n'$. In other words, there are only two permitted configurations, and these are $TE_n \rightarrow TM_n$ and $TM_n \rightarrow TE_n$. The symmetry considerations in cubic ferrites allow for another MO effect, namely the Cotton-Mouton linear birefringence (quadratic in magnetization).

As any three-wave process, a MO interaction can be efficient only under the condition that the interacting waves are in phase synchronism, referred to as the Bragg condition in optics. In the collinear geometry (see Fig. 2), which is better suited for our purposes, and in the TE→TM configuration it reads

$$\Delta\beta = \beta^{(n)TM} - \beta^{(n)TE} = k. \qquad (24)$$

Here *k* is a spin wave wave number. When this condition is satisfied, the integrand in (22) does not contain a rapidly oscillating function, which maximizes the overlap integral. If $\Delta\beta>0$, the scattering MSSW propagates in the same direction as the optical modes. Consequently, the scattered TM-mode is frequency up-shifted $\Omega^{TE} = \Omega^{TM} + \omega$, here $\omega$ is the frequency of the scattering MSW (anti-Stokes process). On the other hand, if $\Delta\beta<0$, the scattering MSW propagates in the opposite direction with respect to the optical modes, as a result the scattered TM-mode is frequency down-shifted $\Omega^{TE} = \Omega^{TM} - \omega$ (Stokes process).

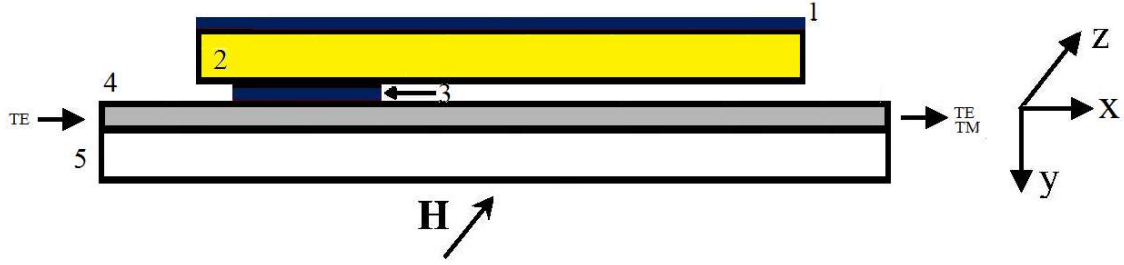

Fig. 2. Geometry of the magnon-light interaction. (1,2,3) is the stripline used to excite magnons in the fiim. (1) stripline ground plane. (2) its dielectric substrate. (3) stripline itself (asymmetric coplanar). (4) YIG film. (5) film substrate. "TE" denotes the non-diffracted light beam and "TM" the diffracted beam (light scattered from magnons). Magnons excited by a microwave current in the stripline propagate in the film collinear with the optical photons. This results in light scattering from the magnons.

In the TM→TE configuration, the Bragg condition reads

$$\Delta\beta = \beta^{(n)TE} - \beta^{(n)TM} = k, \quad (25)$$

and the previous paragraph can be directly applied to the analysis of the scattering mechanism, except that the superscript *TM* should be replaced with *TE* and vice versa.

Now consider the MO coupling coefficient, describing the efficiency of the MO interaction. As mentioned above, in garnets the magnetically induced perturbation of the dielectric permittivity at optical frequencies is comprised of two major contributions due to the Faraday (F) and Cotton-Mouton (CM) effects

$$\Delta\varepsilon_{ij} = \Delta\varepsilon_{ij}^F + \Delta\varepsilon_{ij}^{CM} = if_F \delta_{ijl} M_l + 2g_{44} M_i M_j,$$

where $f_F$ is the Faraday constant, $g_{44}$ is a Cotton-Mouton constant, $M_i$ is the component of the total vector $\mathbf{M} = M\mathbf{e}_z + \mathbf{m}$ along the axis *i*, and summation over the dummy indices is assumed (Einstein convention). Here we have neglected the anisotropic part of the Cotton-Mouton contribution proportional to $g_{11} - g_{12} - 2g_{44}$ which is not essential.

In our case, the total magnetization vector **M** is comprised of the following components

$$\mathbf{M} = M\mathbf{e}_z + \left[ m_x(y)\mathbf{e}_x + m_y(y)\mathbf{e}_y \right] \exp(i\omega t - ikx) \exp(-\nu x)$$

(see (17)).

The amplitude of the scattered optical mode, either TE or TM, i.e. the efficiency of the MO interaction follows from Eq.(22). One finds that it scales as a parameter $\nu_{MO}$ referred to as the index of phase modulation. Its magnitude is given by

$$v_S = 2\pi \frac{\beta_0}{2N_{eff}} \left( i f_F (I_{xz}\tilde{m}_{y,k=\Delta\beta} - I_{yz}\tilde{m}_{x,kk=\Delta\beta}) + 2g_{44}M(I_{xz}\tilde{m}_{x,k=\Delta\beta} + I_{yz}\tilde{m}_{y,k=\Delta\beta}) \right)$$

(26)

for the Stokes process and

$$v_{AS} = 2\pi \frac{\beta_0}{2N_{eff}} \left( f_F (I_{xz}\tilde{m}_{y,k=\Delta\beta} - iI_{yz}\tilde{m}_{x,k=\Delta\beta}) - 2g_{44}M(I_{xz}\tilde{m}_{x,k=\Delta\beta} - iI_{yz}\tilde{m}_{y,k=\Delta\beta}) \right)$$

(27)

for the anti-Stokes one. In Eqs.(26,27), $\beta_0 = 2\pi/\lambda$ is a light wavenumber in vacuum, and $N_{eff}$ is an effective refraction index of the "incident" waveguide mode; $N = N^{TE} = \beta^{(n)TE}/\beta_0$ for the TE→TM optical mode conversion and $N_{eff} = N^{TM} = \beta^{(n)TM}/\beta_0$ for TM→TE. The quantities $\tilde{m}_{x,kk=\Delta\beta}$ and $\tilde{m}_{y,k=\Delta\beta}$ are thickness averages of the respective vector components of the amplitude **m**$_k$ (Eq.(17a)), calculated for $k=\Delta\beta$. Stripline transducers excite spin waves with $kL<1$, therefore the $y$-dependencies of $m_{kx}(y)$ and $m_{ky}(y)$ are close to uniform on the scale of the film thickness. For the purpose of our calculation, it is appropriate to assume that they are perfectly uniform and equal to the thickness averages of these quantities. (The expressions for the latter are given by (A16-A17) in Appendix B.) Under this assumption, Eq.(23) yields $I_{xz} \approx 0$ and $I_{zy} = 1$.

The latter relations illustrate symmetry of the optical waveguide modes. More specifically, the transverse components $E_z^{(n)TE}(y)$ and $E_y^{(n)TM}(y)$ have practically identical spatial distribution across the waveguide and $I_{zy}=1$, while the longitudinal component $E_x^{(n)TM}(y)$ is characterized by a symmetry that is opposite to that of the transverse components, hence $I_{xz} \approx 0$. Thus, both effects, Faraday and Cotton-Mouton, couple only the transverse components of the optical fields, i.e. $E_z^{(n)TE}(y)$ and $E_y^{(n)TM}(y)$.

Taking all this into account, we finally arrive at

$$v_S = \frac{\pi\beta_0}{N_{eff}} \left( -i f_F \tilde{m}_{kx} + 2g_{44}M\tilde{m}_{ky} \right), \qquad (28)$$

$$v_{AS} = \frac{\pi\beta_0}{N_{eff}} \left( i f_F \tilde{m}_{kx} - 2g_{44}M\tilde{m}_{ky} \right). \qquad (29)$$

Here we denoted $\tilde{m}_{x,k=\Delta\beta} \equiv \tilde{m}_{kx}$ and $\tilde{m}_{y,k=\Delta\beta} \equiv \tilde{m}_{ky}$, in order to simplify notations.

The amplitude of the scattered optical mode is proportional to that of the incident one and the coupling coefficient $v_{MO}/2$. Hence the power of the scattered light $I_s$ is given by

$$P_s = \frac{v_{MO}^2}{4} P_L, \qquad (30)$$

where $P_L$ is the power of the incident light and $\nu_{MO} = \nu_S, \nu_{AS}$, depending on whether one deals with a Stokes or an anti-Stokes scattering process. The microwave-photon to optical-photon (microwave-to-light) conversion efficiency is given by Eq.(6) in [7].

$$\eta = \frac{P_s}{P_{in}} \frac{\hbar\omega}{\hbar\Omega}, \quad (31)$$

where $P_{in}$ is given by Eq.(19) and $\Omega$ is the frequency of the light.

### III. Discussion

#### A. Calculation results

In this section, we show results of calculations by employing the theory above. For these calculations, we use the parameters as follows. The thickness of the YIG film $L$=20 micron (Fig. 1). The input transducer is an asymmetric backed coplanar line, having a 400 micron wide signal line and a 200 micron wide ground line ($w_2$=400 micron, $w_3$=200 micron). The gap between the two $\Delta_2$=25 micron. The coplanar line is supported by a 0.5mm-thick dielectric substrate ($d$=0.5mm) with $\varepsilon$=11 whose other surface is metalized and grounded. The length of the transducer $l_a$= 5 mm; it coincides with the width of the film. The applied field $H$=1000 Oe. We use magnetic parameters for the film extracted from a recent microwave experiment at 16 mK [23]: saturation magnetization $4\pi M$=2360 G and gyromagnetic ratio $\gamma$=2.83 MHz/Oe. As the resonance linewidth could not be measured in that experiment (because traveling waves were excited in [23]) we use the fact that no difference between the resonance linewidths at the same temperature and at room temperature was found in [42]. Therefore we use the same magnetic loss parameter as typical for bismuth-substituted YIG films at room temperature: FWHM $2\Delta H$=0.8 Oe.

Bismuth-dopped YIG is the best candidate for the magneto-optical interaction. Unfortunately, its optical constants at 16 mK are not known. However, in [4] it was found that the Verdet constant for a YIG sphere at 16 mK does not differ significantly from the literature value for the room temperature. Therefore, in our calculation we use the literature value for the bismuth-substituted YIG: $f_F$=10$^{-3}$/(4$\pi M$) (measured in G$^{-1}$) obtained for room temperature [43]. The constant $g_{44}$ can then be estimated based on $f_F$ and the Stokes/anti-Stokes peak asymmetry from [18]; one obtains $2g_{44}$=0.58 10$^{-3}$/(4$\pi M$)$^2$ (measured in G$^{-2}$). Optical wavelength is 1.15 µm which corresponds to $\beta_0$= 5.46·10$^6$ rad/m and optical frequency $\Omega$=2$\pi$·2.6 10$^{14}$ Hz. $N$=2, and $P_L$=15 mW as in the recent experiment with a YIG sphere [4].

Let us first calculate the efficiency of the YIG film with asymmetric coplanar transducer as a non-reciprocal device called a microwave isolator. The power transmitted by the device is obtained by calculating the electric field of the spin waves at the position of the output transducer. Then Telegraphist Equations which include a distributed source of a microwave electric voltage are employed, in order to convert the electric field into the output voltage of the output transducer [31]. A result of this calculation is shown in Fig. 3. The distance between the transducers is 4 mm. One sees that transmission characteristics for the +*k* (from Port 1 to Port 2) and –*k* (from Port 2 to Port 1) directions differ by at least 20 dB. Also, the transmission losses in the +*k* directin are just 5 dBm. This characterizes the spin wave device as a very efficient microwave isolator. Note that the magnitude of losses in the +*k* direction is consistent with previous calculations and experiments [28,44].

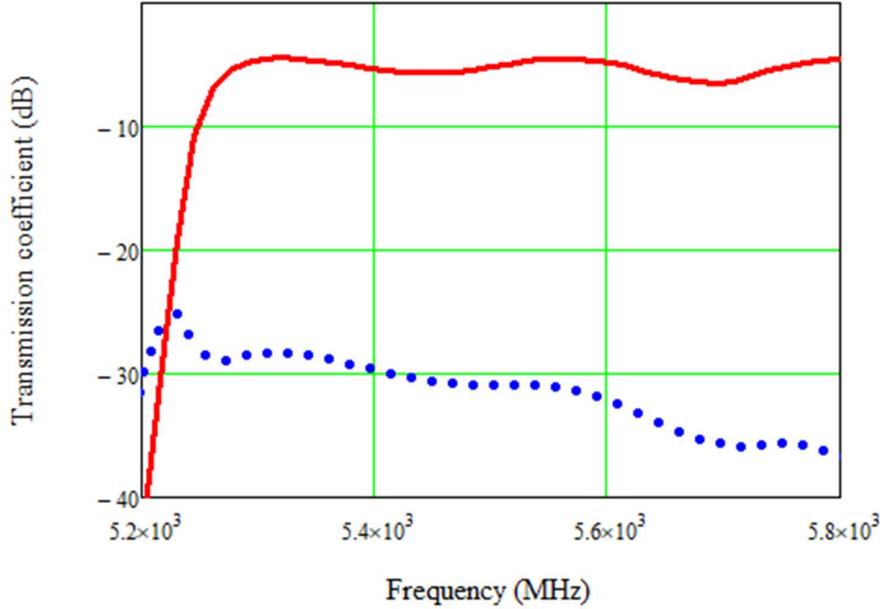

Fig. 3. Non-reciprocity of the spin wave device. Solid line: loss of power transmitted from Port 1 to Port 2 (+*k* direction), dotted line: loss microwave power for transmission from Port 2 to Port 1. Parameters of calculation: *L*=20 micron, $l_a$=5 mm, $w_2$=400 micron, $w_3$=200 micron, $\Delta_2$=25 micron and the distance between the Port 1 and Port 2 transducers is 4 mm. Applied field is 1000 Oe.

Let us now proceed to discussion of the microwave photon to optical photon conversion efficiency $\eta$. Central to the discussion is Eq.(17a), as it enters the expression for $\eta$ (31). One sees that this expression has a resonant denominator – for some $k=k_0$ the real part of the denominator vanishes and the denominator value becomes equal to $i\,\mathrm{Im}[\det(\hat{W}(k))]$. Hence, $\mathrm{Re}[\det(\hat{W}(\Delta\beta))]=0$ corresponds to resonant interaction of light with a spin wave with eigenfrequency $\omega$ given by the condition $\mathrm{Re}[\det(\hat{W}(k=\Delta\beta))]=0$. In the vicinity of $k=\Delta\beta\equiv k_0$ and for small magnetic losses ($\Delta H \ll H$), the denominator can be expanded into Taylor series

$$\det(\hat{W}(k, H + i\Delta H))$$
$$= \mathrm{Re}[\det(\hat{W}(k_0, H)] + (k - k_0)\partial\mathrm{Re}[\det(\hat{W}(k_0, H))]/\partial k + i\Delta H\, \partial\mathrm{Re}[\det(\hat{W}(k_0, H))]/\partial H$$

This expression can be re-written spin-wave frequency resolved:

$$\det(\hat{W}(k, H + \Delta H))$$
$$= \mathrm{Re}[\det(\hat{W}(k_0, H)] + \frac{(\omega - \omega_0)}{V_g(k_0)}\partial\mathrm{Re}[\det(\hat{W}(k_0, H))]/\partial k + i\Delta H\, \partial\mathrm{Re}[\det(\hat{W}(k_0, H))]/\partial H$$

where $\omega_0 = \omega(k_0, H)$ and $\omega(k, H)$ is the dispersion law for spin waves given by $\det(\hat{W}(k)) = 0$. The expression for the spin wave group velocity $V_g$ reads

$$V_g = \frac{d\,\mathrm{Re}\!\left[\det(\hat{W})\right]/dk}{d\,\mathrm{Re}\!\left[\det(\hat{W})\right]/d\omega}.$$ Substituting this expression into the expression above one finds that

the bandwidth of the optical guided mode conversion is given by $\dfrac{\partial\mathrm{Re}[\det(\hat{W})]/\partial\omega}{\partial\mathrm{Re}[\det(\hat{W})]/\partial H}\Delta H$ evaluated at $\omega(k_0), H$. The first term of this expression is close to the gyromagnetic ratio $|\gamma|$ (Eq.(2)). Therefore, the bandwidth of the optical-photon/magnon interaction is given by $\gamma 2\Delta H$, which is the same as in the case of a YIG sphere.

However, for the sphere, the microwave photon to magnon conversion bandwidth is given by the same $\gamma\Delta H$, but for the travelling spin waves the latter bandwidth is much larger – several hundreds of MHz, as one sees from Fig. 3. This implies that multi-channel regime of magnon-optical photon conversion can be implemented, with each individual channel corresponding to a specific pair of optical modes *"n"* with a specific value of TE$_n$-TM$_n$ birefringence $\Delta\beta_n$. (Recall that the $k=\Delta\beta$ condition should be satisfied, therefore the central microwave frequency $\omega_n^{central}(k = \Delta\beta_n)$ for each independent MO channel will correspond to a particular point in the Damon-Eshbach MSW dispersion curve.) The bandwith of each channel, scaling as the inverse of the effective length of the MO interaction $l_{eff}$, typically varies from $10^{-1}$ to 10 MHz. Thus, it is considerably smaller than the above-mentioned frequency band of MSSW excitation by an MSSW transducer.

The important point of the frequency characteristics of a MO channel steming from the Bragg condition based phase synchronism for the optical guided modes and MSSW will be addressed in the last section "B. Ways to further improve the efficiency". In the following, we assume that the phase synchronism Bragg condition $k=\Delta\beta$ has been satisfied at any microwave frequency by properly choosing a respective pair of guided optical modes. Therefore, in all figures below we show the efficiencies of the microwave to optical photon conversion calculated for the maximum of the conversion bandwidth $k=\Delta\beta$. In other words, we assume that the condition $k=\Delta\beta$ is fulfilled for each frequency shown in the graphs. Mathematically this means that the real part of the denominator of Eq.(17a) vanishes for each frequency and $k$ for each frequency is chosen such that it makes $\text{Re}[\det(\hat{W}(k, H + \Delta H))]$ vanish.

Figure 4 shows the result of our calculation for the asymmetric coplanar geometry. One sees that the efficiency of the anti-Stokes process is larger than that of the Stokes one in our geometry. Therefore, below we will focus on the anti-Stokes process.

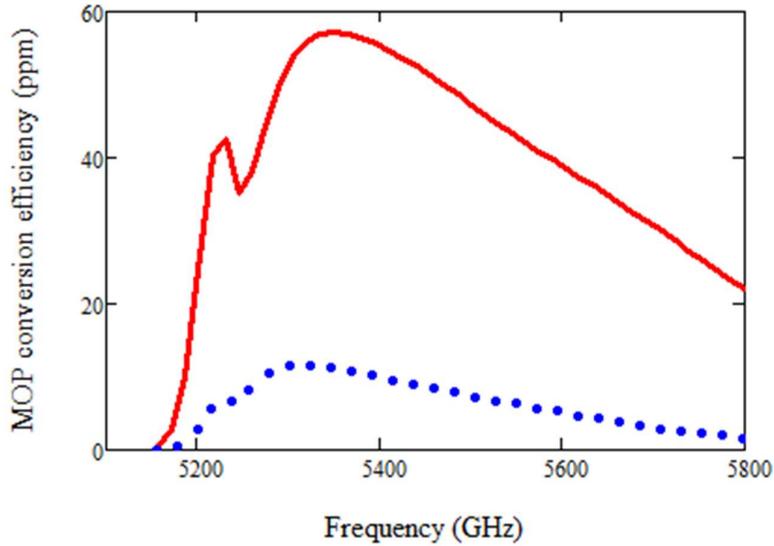

Fig. 4. MOP conversion efficiency for the parameters of Fig. 3 (except the second coplanar transducer is absent, because not needed). Solid line: anti-Stokes process; dotted line: Stokes process. The length of the area where magnons interact with guided optical photons is $3l_f$=20 mm.

In this figure, we actually demonstrate the efficiency of the magnon to optical photon (MOP) conversion. To carry out this calculation, we use the same Eq.(31) but equate $P_{in}$ in it to the power $P^+$ of spin waves radiated by the stripline transducer in the direction of optical mode propagation. The latter is given by Eq.(21).

From the figure, one sees that the light scattering process is characterised by a pronounced maximum. One also sees that the efficiency of MOP conversion is quite high – on the order of $10^{-5}$, or tens of parts per million (PPM). In order to understand formation of the maximum, in Fig. 5 we compare this result to the MOP efficiency in the assumption that a traveling spin wave only exists in a medium. In this case, Eq.(29) can be cast in the form given by Eq.(A18) in Appendix B. This form of the equation for $v_{MO}$ is convenient for calculation of MOP efficiency of eigen-waves. Let us use the case of a film in vacuum ($d=\infty$) as a reference. In Appendix B, we show a simple analytic expression (A20) for the Poynting vector $\Pi_{eig}$ for the eigen-waves in this geometry. By using (A19) and equating $P_{in}$ in (31) to $|\Pi_{eig}|$ one obtains the dotted line in Fig. 5.

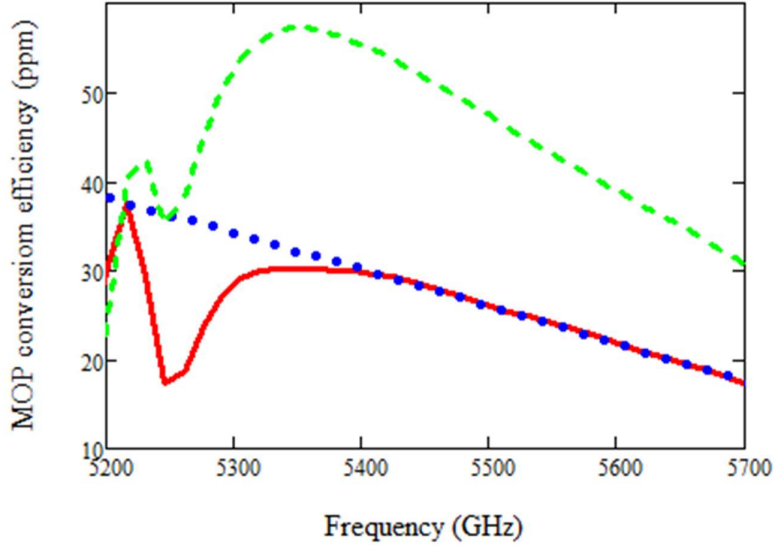

Fig. 5. MOP conversion efficiency for the anti-Stokes process. Dotted line: eigen-waves in a film in vacuum ($d=\infty$). Solid line: Film on top of the backed stripline, but only the traveling spin wave (i.e. far-field) contribution is accounted for. Dashed line: total MOP conversion efficiency for the backed stripline (contributions of both far and near spin wave field of the asymmetric coplanar transducer are taken into account).

The solid line in this figure is obtained by calculating the amplitude of the traveling wave component of the total spin wave field excited by the transducer (in real space, Eq.(17b)). This component and $P^+$ (Eq.(20)) are calculated for $x=0$ and then Eq.(A18) is applied to compute the MOP conversion efficiency. The latter is obtained by equating $P_{in}$ to $P^+(x=0)$ in Eq.(31). One sees that the two curves agree perfectly for frequencies larger than 5.4 GHz. Below this frequency, the efficiency of MOP conversion for the real coplanar device goes down. This is explained by the effect of the metal of the stripline ground plane located at $y=-d$. The effect of the ground plane is present for $kd<1$. The latter wave number range corresponds to the frequency range below 5.4 GHz. When $d$ is increased in the numerical simulation, the maximum in the solid line shifts to lower frequencies. This confirms that the maximum is formed due to the effect of the ground plane.

One also sees that the total MOP conversion efficiency (dashed line) is two times bigger than the traveling spin wave contribution. This evidences that the contribution of the near field of the transducer to the total MOP conversion efficiency may be very significant. Again, the decrease in the efficiency below 5.4 GHz is due to the effect of the metal ground plane.

The respective total microwave photon to optical photon (MPOP) efficiency $\eta$ (Eq.(31)) and the efficiency of microwave photon to magnon conversion $N_m$ (Eq.(21)) is shown in Fig. 6. One sees that the MPOP efficiency is of the same order of magnitude as the MOP one. This is because the efficiency of the microwave photon to magnon conversion $N_m$ by the asymmetric coplanar transducer is close to 1.

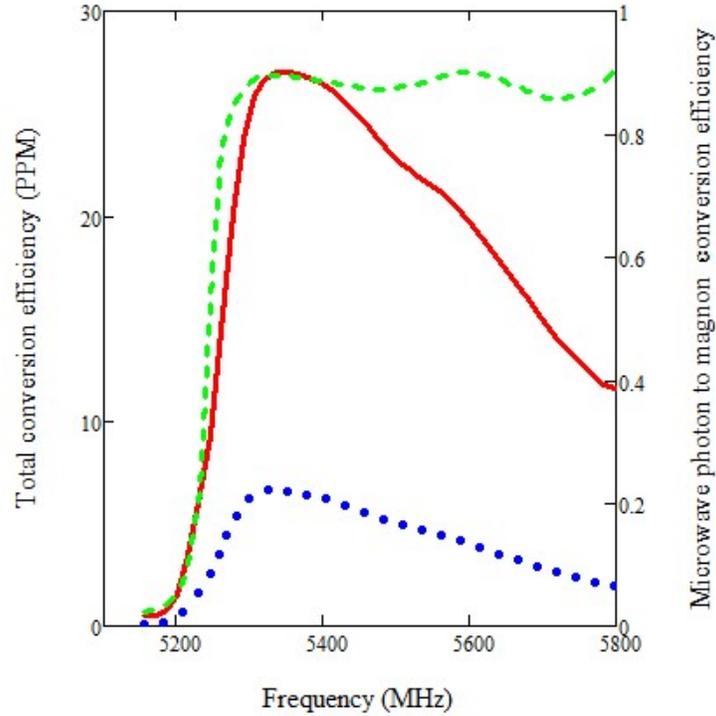

Fig. 6. Total (MPOP) conversion efficiency (left-hand axis) for the asymmetric coplanar line geometry from Fig. 5. Solid line: anti-Stokes process; dotted line (Stokes process). Dashed line: microwave photon to magnon conversion efficiency by the asymmetric coplanar transducer (right-hand axis).

### B. Ways to further improve the efficiency

Even though MPOP efficiency achievable with travelling magnons is significantly larger than one achievable with a standing wave oscillation [4], it is still much smaller than for optomechanical converters (10%). Therefore, below we discuss ways to further improve the conversion efficiency. However, one has to keep in mind that all these measures will decrease the available frequency bandwidth.

The first obvious way to increase efficiency is by decreasing the area of cross-section for light scattering from magnons. Figure 7 demonstrates the results of calculations of MOP conversion efficiency for three different film cross-sections. The first one is the same as in Fig. 4 – $L=20$ micron, $l_a=5$ mm. The second one is for the film thickness $L=4$ micron, which is the same thickness as in the experiments [18,19], but keeping the film width the same $l_a=5$ mm. The third calculation is for a microscopic film cross-section $L=4$ micron, $l_a=50$ micron. One sees that the decrease in the film thickness alone does not change the efficiency; this is because, ultimately, the efficiency scales as $kL$, as does the frequency of the Damon-Eshbach waves for $kL<<1$. In other words, $k$-vector resolved, one gains in efficiency by decreasing $L$. However, frequency-resolved, the net gain is zero, because the spin wave frequency for a given $k$ decreases proportionally, such that for the same frequency the efficiency remains the same. This is illustrated in Fig. 7 by showing $k(\omega)$ dependences for both film thicknesses.

On the other hand, the decrease in the film width has an impact on the efficiency, as the same figure shows. Here we have to note that for the 4x50 square micron cross-section, $L$ and $l_a$ are comparable, therefore, strictly speaking, our theory developed under assumption $L<< l_a$ is not fully applicable in this case. Notwithstanding this, it demonstrates that the reduction of the film width is a valid way to significantly boost the efficiency of the MOP conversion. However, the decrease in $l_a$ decreases coupling of the microwave field of the transducer to the precessing spins. This happens because the number of spins in the volume where the microwave field is present shrinks with the decrease in $l_a$. In our case of a stripline shorted at its end this is seen as a decrease in the input antenna impedance for $l_a$ =50 micron. It is possible to compensate the drop in the impedance to a significant extent by playing around with the geometry of the coplanar line, as shown in Fig. 8. To this end, one has to use a coplanar line with a microscopic cross-section. As this calculation shows, in this way it is possible to achieve an impressive total conversion efficiency of 0.3 percent. Importantly, the difference between the MPOP and MOP efficiencies is about 3 times in this case, which is not a huge number, and the input impedance of the coplanar transducer is large enough – about 8 Ohm. Hence, it should be technically easy to increase MPOP to 0.9% by inserting an impedance matching circuit between the feeding line and the coplanar transducer.

The conversion efficiency of almost 1% is a very good result. However, the possibility of its further improvement through additional reduction of the YIG film width should be taken with caution, as our theory is not fully applicable for $l_a$ comparable to $L$.

One more way to improve the efficiency is by further increasing the time of spin wave interaction with the optical modes. This can be done by confining the optical field inside a length equal to $l_f$ by forming an optical resonator of this length. Then the time of spin wave interaction with the optical modes will increase by $Q$ times, where $Q$ is the quality factor for the optical resonator.

To be specific, let us begin with a modest and easily realizable quality factor of $Q = 100$. Naturally, this will increase the length of magnon-light interaction in a 4 μm thick YIG film from 0.6 cm to 0.6 m which will increase the efficiency accordingly from $10^{-2}$ to 1 (or 100%), as the efficiency scales as $l_f$ (see Eq.(A18)). At the same time, this increase in the interaction length will inevitably lead to restrictions imposed by the Bragg selectivity in the reciprocal space. As a result, even a slight deviation from the phase-synchronism Bragg condition will lead to a drop in the efficiency of the MO interaction according to $\text{sinc}^2(\Delta\varphi)$ where $\Delta\varphi$ is the phase mismatch due to the above-mentioned deviation. To be specific, let us consider the TE→TM configuration with $\Delta\beta > 0$ (see Eq.(24)), in which case the magnon propagating in the same direction as the "incident" TE optical mode will contribute to the anti-Stokes process (the up-shifted scattered TM mode). There are two mechanisms, both dispersion related, contributing to this mismatch: magnetic (MSSW mode) and optical (optical waveguide mode).

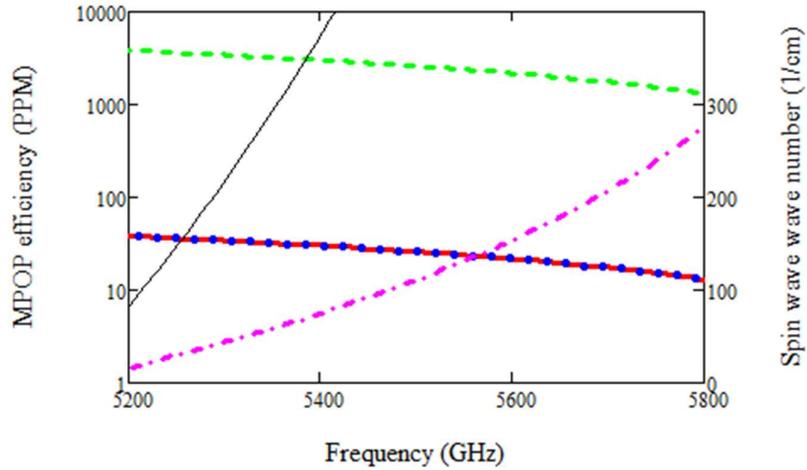

Fig. 7. Thick solid line: conversion efficiency for $L$=20 micron, $l_a$=5 mm; dotted line: the same, but $L$=4 micron, $l_a$=5 mm; dashed line: the same, but $L$=4 micron, $l_a$=50 micron. Thin solid line: spin wave wave number for $L$=4 micron (right-hand axis); dash-dotted line: the same, but for $L$=20 micron.

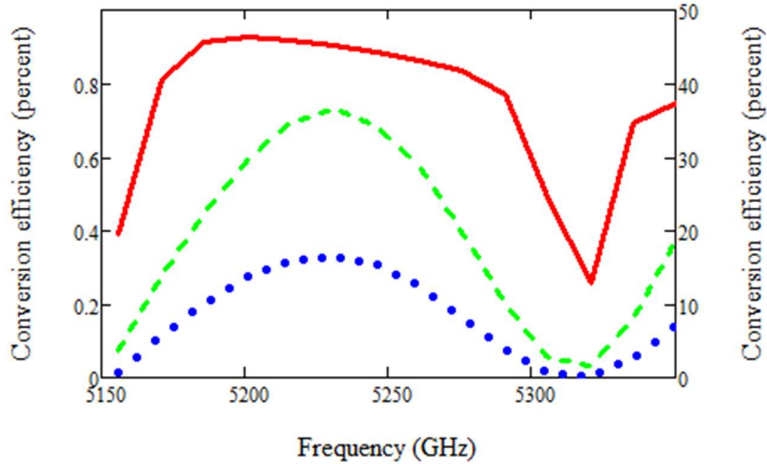

Fig. 8. Solid line: MOP conversion efficiency or the microscopic device (left-hand axis). Dotted line: total (MPOP) conversion efficiency (left-hand axis). Dashed line (right-hand axis): microwave photon to magnon conversion efficiency by the asymmetric coplanar transducer ($w_1=\Delta_1=0$). Film thickness: $L$=4 micron, film width: 50 micron. Width of the signal line of the asymmetric coplanar transducer: $w_2$=4 micron, width of the ground line: $w_3$=20 micron, gap between the two: $\Delta_2$=250 micron. Distance to the ground plane $d$=500 micron. Dielectrip permittivity of the substrate of the coplanar line: 11.

First, numerical estimation of the value of the wave number of the scattered optical mode should take into account the Doppler shift imposed by the moving magnon, i.e. $\beta^{TM}(\Omega+\omega) \approx \beta^{TM}(\Omega) + \omega/V_g^{opt}$, where $V_g^{opt}$ is the group velocity of the optical mode. The latter approximation is perfectly justified since $\omega$ is really very small with respect to $\Omega$ and typically $\omega/\Omega \sim 10^{-5}$. Thus, the Doppler related additional phase shift reads $\Delta\varphi \approx \delta\beta\, l_{eff}/2 = \omega l_{eff}/(2V_g^{opt})$, here $l_{eff}$ is the effective interaction length, while $\delta\beta = \delta\beta(\omega)$ is the perturbation of the wavenumber of the scattered optical TM mode due to the Doppler frequency shift and, consequently, $\delta\beta(0) = 0$. The optical wave number frequency dependence $\delta\beta(\omega)$ makes the total phase mismatch also frequency dependent $\Delta\varphi = \Delta\varphi(\omega)$. Suppose that in the absence of the Doppler shift the Bragg condition is fully satisfied and the phase synchronism is perfect, i.e. $\Delta\varphi(0) = 0$. Now let us estimate the consequences of the Doppler effect, namely the deviation from the phase synchronism $\delta\varphi(\omega) = \Delta\varphi(\omega) - \Delta\varphi(0)$ for a MSSW frequency of 4 GHz and two characteristic values of the effective interaction length mentioned above $l_{eff} = 0.6$ m (with the optical resonator) and $l_{eff} = 6$ mm (without the resonator). In the conventional no-resonance case one obtains $\Delta\varphi(4\text{GHz}, 6\text{mm}) \approx 0.5\,\text{rad}$ which is not enough to influence appreciably the mechanism of the MO interaction that is why this effect was neglected in the earlier papers of the 1980s – 1990s. It follows from the same analysis that the effective bandwidth of the MO interaction defined as full width at half maximum (FWHM) and adapted for the *Sinc* function is equal to $\Delta f_{MSSW}(6\text{mm}) \approx 30\,\text{GHz}$. In the resonator case one obtains an impressive figure of $\Delta\varphi(\omega) \approx 0.5 \cdot 10^2\,\text{rad}$ and, as a result, $\Delta f_{MSSW}(0.6\text{m}) \approx 300\,M\text{Hz}$.

Second, the MSSW dispersion also appears in the Bragg condition. Moreover, the phase synchronism is even more sensitive to MSSW frequency variations through this mechanism which is due to the direct presence of the wavenumber of extremely slow MSSW modes in the Bragg condition. As in the previous case, the coefficient weighing this contribution is the inverse group velocity, but this time that of the MSSW that is about $10^4$ times slower. Let us suppose that the 4µm thick YIG film is magnetized to saturation by a 800 Oe magnetic field. In this case a MSSW with a frequency of 4 GHz will propagate with a group velocity $V_g$ of approximately $4 \cdot 10^4$ m/s. Correspondingly, in the "without-resonator" and with-resonator configurations one obtains $\Delta f_{MSSW}(6\text{mm}) \approx 8\,\text{MHz}$ and $\Delta f_{MSSW}(0.6\text{m}) \approx 80\,\text{KHz}$ respectively. The former figure was confirmed in the experiments in the 1980s-1990s. In any case, it is the second mechanism that bottlenecks the frequency properties, its bandwidth in the first place, of the MO interaction.

Thus, in this paragraph we provide useful quantitative data on the « bandwidth – efficiency » limitations, classical in photonics, which stem from the general laws of three-wave interactions requiring phase synchronism between the interacting waves. This information is indispensable for the design of specific devices specialized in coherently connecting distant superconducting qubits via light.

On the other hand, our analysis emphasizes the importance of the role played by the Doppler shift in the scattering of light by MSSW in the case where the effective interaction length $l_{eff}$ exceeds a critical value of several centermeters in the lower GHz band addressed in this paper. While it is of less importance in the case of Brillouin light scattering by thermally excited incoherent magnons [5], it should not be overlooked when one considers coherent magnon-qubit up conversion to optical frequencies [45]. It is especially important for MO interactions involving the homogeneous Kittel mode in which case $\Delta\beta \to 0$, especially if the optical resonator quality factor is as high as $10^5$ [46]. In other words, the actual pertinent criterion of the smallness of the Doppler shift [45] cannot be formulated solely in terms of the ratio of the frequencies of the interacting waves $\frac{\omega}{\Omega} \ll 1$, even if it is as small as $10^{-5}$. It must follow directly from the Bragg phase synchronism and be expressed in terms of the maximum tolerable phase mismatch, thus reading

$$\Delta\varphi = \frac{\omega}{v_{gr}^{opt}} \frac{l_{eff}}{2} < 1.$$

It is should emphasized that boosting the efficiency of the MO interaction through a radical increase of the interaction length cannot be implemented in the absence of a reliable mechanism of fine-tuning to the Bragg condition. In this regard, the configuration relying on travelling spin wave has an advantage of an additional flexible degree of freedom, namely the tunable spatial periodicity in the form of the spin wave wavenumber $k$.

Another important aspect of the Doppler frequency shift is its asymmetry with respect to the inversion of the direction of the incident optical mode which can be exploited in order to create non-reciprocal MO devices. Thus, if both interacting waves, the spin wave and the incident optical mode, propagate collinearly this shift will be positive ($+\omega$), whereas anti-collinear propagation will produce a negative shift ($-\omega$). This, in its turn, means that if the Bragg condition is perfectly satisfied in the collinear geometry $\Delta\varphi(+\omega) = 0$, reversal of propagation direction of the optical incident wave will lead to a double phase asynchronism $\Delta\varphi(-\omega) = 2\frac{\omega}{v_{gr}^{opt}} \frac{l_{eff}}{2}$. As a result, the amplitude of the "new-born" scattered optical wave will reduce accordingly. In other words, such a MO element can be regarded as an optical isolator with a ratio of nonreciprocity (which is defined as the ratio of amplitudes of counter propagating optical modes) equal to $Sinc\ (\Delta\varphi(-\omega))$.

## IV. Conclusions

In this work we evaluated theoretically the efficiencies of a travelling magnon based microwave to optical photon converter for applications in Quantum Information. The microwave to optical photon conversion efficiency was found to be larger than in a similar process employing a YIG sphere by at least 4 orders of magnitude. By employing an optical resonator of a large length (such that the traveling magnon decays before forming a standing wave over the resonator length) it will be possible to further increase the efficiency by several orders of magnitude, potentially reaching a magnitude similar to one achieved with opto-mechanical resonators. However, this measure will decrease the frequency bandwidth of conversion.

Also, as a spin-off result, it has been shown that microwave isolation of more that 20 dB with direct insertion loss of about 5 dBm can be achieved with YIG film based isolators. These devices are needed to isolate qubits from noise in a microwave circuit to which they are connected.

An important advantage of the concept of the travelling spin wave based Quantum Information devices is a perfectly planar geometry and a possibility of implementing a device as a hybrid opto-microwave chip.

**Acknowledgement**

Research Colaboration Award from the University of Western Australia is acknowledged. The authors also thank M. Goryachev, M. Tobar and V.N. Malyshev for fruitful discussions.

**Appendix A: Solution of the system of equations (2-5).**

The solution is obtained in the Fourier space (Eq.(6)).

$$m_{kx} = A\exp(|k|y) + B\exp(-|k|y), \quad (A1)$$

$$m_{ky} = AC_{my}(|k|)\exp(|k|y) + BC_{my}(-|k|)\exp(-|k|y) \quad (A2)$$

where

$$C_{my}(q) = i\frac{\omega_M qk - \omega(q^2 - k^2)}{\omega_H q^2 + (q^2\omega_M - k^2\omega_H)}. \quad (A3)$$

Similarly,

$$h_{kx} = AC_{hx}(|k|)\exp(|k|y) + B(C_{hx}(-|k|))\exp(-|k|y), \quad (A4)$$

$$h_{ky} = AC_{hy}(|k|)\exp(|k|y) + B(C_{hy}(-|k|))\exp(-|k|y), \quad (A5)$$

with

$$C_{hx}(q) = \frac{k(\omega q + k\omega_H)}{\omega_H q^2 + (q^2\omega_M - k^2\omega_H)}, \quad (A6)$$

and

$$C_{hy}(q) = i\frac{q(\omega q + k\omega_H)}{\omega_H q^2 + (q^2\omega_M - k^2\omega_H)}, \quad (A7)$$

where $\omega_H = \gamma H$ and $\omega_M = \gamma M$ (or $\omega_M = \gamma 4\pi M$ in Gaussian units).

Application of the electro-dynamic boundary conditions results in a vector-matrix equation

$$\hat{W}\begin{bmatrix} A \\ B \end{bmatrix} = \begin{bmatrix} i\,\text{sign}(k)j_k \\ 0 \end{bmatrix}, \quad (A8)$$

where

$$\hat{W} = \begin{bmatrix} \alpha(|k|) & \alpha(-|k|) \\ \beta(|k|) & \beta(-|k|) \end{bmatrix}, \quad (A9)$$

$$\alpha(q) = \left[ C_{my}(q) + C_{hy}(q) \right] \coth(|q|d) - i\,\text{sign}(q) C_{hx}(q), \quad (A10)$$

and

$$\beta(q) = \left[ C_{my}(q) + C_{hy}(q) + i\,\text{sign}(q) C_{hx}(q) \right] \exp(qL). \quad (A11)$$

Solving (A8) with respect to the vector on its left-hand side yields

$$\begin{bmatrix} A \\ B \end{bmatrix} = \hat{W}^{-1} \begin{bmatrix} i\,\text{sign}(k)j_k \\ 0 \end{bmatrix}, \quad (A12)$$

where

$$\hat{W}^{-1} = \frac{1}{\det(\hat{W})} \begin{bmatrix} \beta(-|k|) & -\alpha(-|k|) \\ -\beta(|k|) & \alpha(|k|) \end{bmatrix}, \quad (A13)$$

and $\det(\hat{W})$ denotes the determinant of the matrix $\hat{W}$.

Accordingly,

$$A = i\,\text{sign}(k)\beta(|k|)j_k / \det(\hat{W}), \quad (A14)$$

$$B = -i\,\text{sign}(k)\beta(-|k|)j_k / \det(\hat{W}). \quad (A15)$$

This concludes the solution of the problem of calculation of the spin wave amplitude $\mathbf{m}_k$. The closed form of the expression for $\mathbf{m}_k$ is given by Eq.(17(a)).

**Appendix B: Calculation of quantities entering expressions for magnon to optical photon conversion efficiency**

The thickness-averaged spin wave amplitude is obtained from Eqs. (A1) and (A2) and reads:

$$\tilde{m}_{kx} = A\,F(|k|) + B\,F(-|k|), \quad \text{(A16)}$$

$$\tilde{m}_{ky} = A\,C_{my}(|k|)\,F(|k|) + B\,C_{my}(-|k|)\,F(-|k|), \quad \text{(A17)}$$

where $F(k) = \mathrm{sign}(k)(\exp(kL)-1)/(kL)$.

For eigen-waves we may set $A=1$. Then (A14) and (A15) yield $B = -\beta(-|k|)/\beta(|k|)$. Substituting these $A$ and $B$ into (A16) and (A17) and the result into (28) and (29) we obtain the MOP conversion efficiency. In doing this we need to specify the length of the MO interaction area. It is found introducing the spin wave propagation path as $l_f = \gamma \Delta H / V_g$, where $\Delta H$ is the magnetic loss parameter and $V_g = \partial \omega / \partial k$ is the group velocity of spin waves. Then the wave decays exponentially during its propagation $\tilde{\mathbf{m}}_k = \tilde{\mathbf{m}}_k^0 \exp(-x/l_f)\exp(-ikx)$, where $\tilde{\mathbf{m}}_k^0$ is its initial amplitude (at $x=0$). Substituting this expression into (17b) and it into the overlap integral (22) we arrive at the expression for the light-magnon coupling coefficient for the eigen-waves.

$$v_{AS} = \frac{\beta_0 l_f}{2N}\left(f_F(I_{xz}\tilde{m}_y^0 - iI_{yz}\tilde{m}_x^0) - 2g_{44}M(I_{xz}\tilde{m}_x^0 - iI_{yz}\tilde{m}_y^0)\right). \quad \text{(A18)}$$

In order to calculate the MOP conversion efficiency for the eigenwaves we also need the Poynting vector for them. It is obtained from (A2), (A4), (A5) and (5). For a film in vacuum (no metal ground plane at $y=-d$) it reads:

$$\Pi(x) = \frac{\omega\mu_0 l_s}{2k}\mathrm{Re}(\Pi_1 + \Pi_2 + \Pi_3)\exp(-2x/l_f), \quad \text{(A19)}$$

where

$\Pi_1 = |A_1|^2/(2|k|)$, $\Pi_3 = |B_3|^2 \exp(-2|k|L)/(2|k|)$, $A_1 = AC_{hx}(|k|) + BC_{hx}(-|k|)$,
$B_3 = AC_{hx}(|k|)\exp(2|k|L) + BC_{hx}(-|k|)$,

and

$$\Pi_2 = A\overline{B}L\left[C_{hy}(|k|) + C_{my}(|k|)\right]\overline{C}_{hy}(-|k|) + \overline{A}BL\left[C_{hy}(-|k|) + C_{my}(-|k|)\right]\overline{C}_{hy}(|k|) + $$
$$+ |A|^2\left[C_{hy}(|k|) + C_{my}(|k|)\right]\overline{C}_{hy}(|k|)\left[1-\exp(2|k|L)\right]/(2|k|)$$
$$+ |B|^2\left[C_{hy}(-|k|) + C_{my}(-|k|)\right]\overline{C}_{hy}(-|k|)\left[1-\exp(-2|k|L)\right]/(2|k|)$$
.

The MOP conversion efficiency is obtained by substituting (A19) into (31) and using $\Pi(x=0)$ (Eq.(A19)) as $P_{\text{in}}$.